\documentclass[pdftex,epjc3, twocolumn]{svjour3}
\journalname{Eur. Phys. J. C}

\RequirePackage{mathptmx}      % use Times fonts if available on your TeX system
\RequirePackage[numbers,sort&compress]{natbib}

\usepackage{amsmath}
\usepackage{amssymb}
\usepackage{dcolumn}% Align table columns on decimal point
\usepackage{bm}% bold math
\usepackage{tabularx}
\usepackage{booktabs}
\usepackage{xspace}
\usepackage{xcolor}
\usepackage{appendix}
\usepackage{adjustbox}
\usepackage[caption=false]{subfig}%
\usepackage{textcomp}
\usepackage{upgreek} 
\usepackage[switch]{lineno}     % Line number

\renewcommand{\vec}[1]{\boldsymbol{#1}}
\newcommand{\cSone}{\ensuremath{\mathrm{cS1}}\xspace}
\newcommand{\cStwo}{\ensuremath{\mathrm{cS2}}\xspace}
\newcommand{\cStwob}{\ensuremath{\mathrm{cS2_b}}\xspace}

\newcommand{\Erec}{\ensuremath{E_\mathrm{rec}}\xspace}
\newcommand{\Etrue}{\ensuremath{E_\mathrm{true}}\xspace}
\newcommand{\Erecperp}{\ensuremath{E_\mathrm{rec}^\perp\xspace}}

\newcommand{\radius}{R}

\newcommand{\micros}{\ensuremath{\upmu\mathrm{s}}}
\renewcommand{\L}{\ensuremath{\mathcal{L}}}
\newcommand{\migmat}{\ensuremath{\mathcal{P}}}
\newcommand{\radonttz}{\ensuremath{{}^{220}\mathrm{Rn} }}

\newcommand{\given}{\ensuremath{\mid}}

\title{An approximate likelihood for nuclear recoil searches with XENON1T data}

\author{E.~Aprile\thanksref{addr3}
\and
K.~Abe\thanksref{addr23}
\and
F.~Agostini\thanksref{addr0}
\and
S.~Ahmed Maouloud\thanksref{addr18}
\and
M.~Alfonsi\thanksref{addr5}
\and
L.~Althueser\thanksref{addr7}
\and
B.~Andrieu\thanksref{addr18}
\and
E.~Angelino\thanksref{addr14}
\and
J.~R.~Angevaare\thanksref{addr8}
\and
V.~C.~Antochi\thanksref{addr12}
\and
D.~Ant\'on Martin\thanksref{addr1}
\and
F.~Arneodo\thanksref{addr9}
\and
L.~Baudis\thanksref{addr17}
\and
A.~L.~Baxter\thanksref{addr10}
\and
L.~Bellagamba\thanksref{addr0}
\and
R.~Biondi\thanksref{addr4}
\and
A.~Bismark\thanksref{addr17}
\and
A.~Brown\thanksref{addr19}
\and
S.~Bruenner\thanksref{addr8}
\and
G.~Bruno\thanksref{addr9,addr13}
\and
R.~Budnik\thanksref{addr16}
\and
C.~Capelli\thanksref{addr17}
\and
J.~M.~R.~Cardoso\thanksref{addr2}
\and
D.~Cichon\thanksref{addr6}
\and
B.~Cimmino\thanksref{addr20}
\and
M.~Clark\thanksref{addr10}
\and
A.~P.~Colijn\thanksref{addr8}
\and
J.~Conrad\thanksref{addr12}
\and
J.~J.~Cuenca-Garc\'ia\thanksref{addr26}
\and
J.~P.~Cussonneau\thanksref{addr13}
% ...
(XENON Collaboration\thanksref{email1}). }
\newcommand{\bologna}{Department of Physics and Astronomy, University of Bologna and INFN-Bologna, 40126 Bologna, Italy}
\newcommand{\chicago}{Department of Physics \& Kavli Institute for Cosmological Physics, University of Chicago, Chicago, IL 60637, USA}
\newcommand{\coimbra}{LIBPhys, Department of Physics, University of Coimbra, 3004-516 Coimbra, Portugal}
\newcommand{\columbia}{Physics Department, Columbia University, New York, NY 10027, USA}
\newcommand{\lngs}{INFN-Laboratori Nazionali del Gran Sasso and Gran Sasso Science Institute, 67100 L'Aquila, Italy}
\newcommand{\mainz}{Institut f\"ur Physik \& Exzellenzcluster PRISMA$^{+}$, Johannes Gutenberg-Universit\"at Mainz, 55099 Mainz, Germany}
\newcommand{\heidelberg}{Max-Planck-Institut f\"ur Kernphysik, 69117 Heidelberg, Germany}
\newcommand{\munster}{Institut f\"ur Kernphysik, Westf\"alische Wilhelms-Universit\"at M\"unster, 48149 M\"unster, Germany}
\newcommand{\nikhef}{Nikhef and the University of Amsterdam, Science Park, 1098XG Amsterdam, Netherlands}
\newcommand{\nyuad}{New York University Abu Dhabi - Center for Astro, Particle and Planetary Physics, Abu Dhabi, United Arab Emirates}
\newcommand{\purdue}{Department of Physics and Astronomy, Purdue University, West Lafayette, IN 47907, USA}
\newcommand{\rice}{Department of Physics and Astronomy, Rice University, Houston, TX 77005, USA}
\newcommand{\stockholm}{Oskar Klein Centre, Department of Physics, Stockholm University, AlbaNova, Stockholm SE-10691, Sweden}
\newcommand{\subatech}{SUBATECH, IMT Atlantique, CNRS/IN2P3, Universit\'e de Nantes, Nantes 44307, France}
\newcommand{\torino}{INAF-Astrophysical Observatory of Torino, Department of Physics, University  of  Torino and  INFN-Torino,  10125  Torino,  Italy}
\newcommand{\ucsd}{Department of Physics, University of California San Diego, La Jolla, CA 92093, USA}
\newcommand{\wis}{Department of Particle Physics and Astrophysics, Weizmann Institute of Science, Rehovot 7610001, Israel}
\newcommand{\zurich}{Physik-Institut, University of Z\"urich, 8057  Z\"urich, Switzerland}
\newcommand{\paris}{LPNHE, Sorbonne Universit\'{e}, CNRS/IN2P3, 75005 Paris, France}
\newcommand{\freiburg}{Physikalisches Institut, Universit\"at Freiburg, 79104 Freiburg, Germany}
\newcommand{\napels}{Department of Physics ``Ettore Pancini'', University of Napoli and INFN-Napoli, 80126 Napoli, Italy}
\newcommand{\nagoya}{Kobayashi-Maskawa Institute for the Origin of Particles and the Universe, and Institute for Space-Earth Environmental Research, Nagoya University, Furo-cho, Chikusa-ku, Nagoya, Aichi 464-8602, Japan}
\newcommand{\laquila}{Department of Physics and Chemistry, University of L'Aquila, 67100 L'Aquila, Italy}
\newcommand{\tokyo}{Kamioka Observatory, Institute for Cosmic Ray Research, and Kavli Institute for the Physics and Mathematics of the Universe (WPI), University of Tokyo, Higashi-Mozumi, Kamioka, Hida, Gifu 506-1205, Japan}
\newcommand{\kobe}{Department of Physics, Kobe University, Kobe, Hyogo 657-8501, Japan}

\newcommand{\kit}{Institute for Astroparticle Physics, Karlsruhe Institute of Technology, 76021 Karlsruhe, Germany}
\newcommand{\tsinghua}{Department of Physics \& Center for High Energy Physics, Tsinghua University, Beijing 100084, China}
\newcommand{\alsoatferrara}{INFN, Sez. di Ferrara and Dip. di Fisica e Scienze della Terra, Universit\`a di Ferrara, via G. Saragat 1, Edificio C, I-44122 Ferrara (FE), Italy}

\newcommand{\alsoatcoimbrapoli}{Coimbra Polytechnic - ISEC, 3030-199 Coimbra, Portugal}
\newcommand{\alsoatiarnagoya}{Institute for Advanced Research, Nagoya University, Nagoya, Aichi 464-8601, Japan}
\authorrunning{XENON Collaboration}
\thankstext{addr31}{Also at \alsoatiarnagoya}
\thankstext{addr30}{Also at \alsoatcoimbrapoli}
\thankstext{addr28}{Also at \alsoatferrara}

\thankstext{emailh}{\texttt{hagar.landsman@weizmann.ac.il}}
\thankstext{emailk}{\texttt{knut.dundas.moraa@columbia.edu}}
\thankstext{emailj}{\texttt{jpienaar@uchicago.edu}}
\thankstext{email1}{\texttt{xenon@lngs.infn.it}}
\institute{\columbia\label{addr3}
\and
\tokyo\label{addr23}
\and
\bologna\label{addr0}
\and
\paris\label{addr18}
\and
\mainz\label{addr5}
\and
\munster\label{addr7}
\and
\torino\label{addr14}
\and
\nikhef\label{addr8}
\and
\stockholm\label{addr12}
\and
\chicago\label{addr1}
\and
\nyuad\label{addr9}
\and
\zurich\label{addr17}
\and
\purdue\label{addr10}
\and
\lngs\label{addr4}
\and
\freiburg\label{addr19}
\and
\subatech\label{addr13}
\and
\wis\label{addr16}
\and
\coimbra\label{addr2}
\and
\heidelberg\label{addr6}
\and
\napels\label{addr20}
\and
\kit\label{addr26}
\and
\laquila\label{addr22}
\and
\rice\label{addr11}
\and
\tsinghua\label{addr27}
\and
\nagoya\label{addr21}
\and
\ucsd\label{addr15}
\and
\kobe\label{addr24}
}
\begin{document}
\date{}
\maketitle
\abstract{
The XENON collaboration has published stringent limits on specific dark matter -nucleon recoil spectra from dark matter recoiling on the liquid xenon detector target. In this paper, we present an approximate likelihood for the XENON1T 1\,tonne-year nuclear recoil search applicable to any nuclear recoil spectrum. Alongside this paper, we publish data and code to compute upper limits using the method we present. 
The approximate likelihood is constructed in bins of reconstructed energy, profiled along the signal expectation in each bin. This approach can be used to compute an approximate likelihood and therefore most statistical results for any nuclear recoil spectrum.
Computing approximate results with this method is approximately three orders of magnitude faster than the likelihood used in the original publications of XENON1T, where limits were set for specific families of recoil spectra. 
Using this same method, we include toy Monte Carlo simulation-derived binwise likelihoods for the upcoming XENONnT experiment that can similarly be used to assess the sensitivity to arbitrary nuclear recoil signatures in its eventual 20\,tonne-year exposure.
\nocite{xenon1t_binwise_data}
}

\maketitle

\section{Introduction}
Persuasive astrophysical and cosmological evidence for the existence of dark matter has led to numerous direct detection efforts for weakly interacting massive particles (WIMPs) over the last 20 years~\cite{Schumann:2019eaa}. Amongst these was the \\XENON1T~\cite{xenon1t_instrument} experiment, which collected 1\,tonne-year of exposure from 2016 to 2018. It culminated in the then most stringent limits for spin-independent (SI) WIMP nucleon interactions above 6 GeV/c$^2$~\cite{xenon1t_sr1} at the time. Subsequent limits on spin-dependent WIMP interactions with neutrons and protons~\cite{xenon1t_SD} as well as WIMP-pion couplings~\cite{xenon1t_pion} have been published. In each of the aforementioned interactions the expected signature in the detector is a nuclear recoil (NR), induced by the single scatter of a WIMP off a xenon atom. All four searches use the same background and detector response models and NR search data set.  Other XENON results may be applicable to some NR interactions, such as ionisation-only signatures~\cite{xenon1t_s2only} or Migdal effect searches~\cite{xenon1t_migdal}. For WIMPs below $\sim10\mathrm{GeV}/c^2$, the best XENON1T limits are provided by a dedicated low-energy NR search searching for solar ${}^8$B neutrinos~\cite{xenon1t_cevns}.
The SI recoil spectrum and a fixed halo model are the standard for reporting direct-detection WIMP searches~\cite{resultreporting}.  Different interactions or dark matter fluxes, either from alternate dark matter halo models~\cite{Evans:2018bqy} or methods for generating boosted dark matter~\cite{McKeen:2018pbb} can yield different spectra. As the exact halo parameters are uncertain, and any candidate dark matter particle may interact through a number of different channels, a robust method to constrain arbitrary nuclear recoil spectra is required.

In the full likelihood used for the XENON1T NR \\searches there are two data-taking periods, each with an accompanying electronic recoil (ER) calibration set and ancillary measurement terms constraining the detector response and microphysics parameters as well as background models represented in 20 nuisance parameters~\cite{xenon1t_analysis2}.
Each science data set is modelled in  
three analysis dimensions (discussed in Section~\ref{sec:dimensions}), with five background components (presented in~\ref{sec:xenon1t}).
This complexity was reflected in the computational expense, requiring about $\sim$\,30\,s for a toy Monte Carlo (toy-MC) simulation of the analysis.

In this paper we present the profiled likelihood of the XENON1T NR search in bins of reconstructed energy, a description of how it may be used to calculate upper limits for a generic NR spectrum and a data release with accompanying code~\cite{xenon1t_binwise_data} allowing the physics community to use this method to recast the XENON1T result. The computation is fast, taking about $\sim$\,40\,ms to compute an upper limit for a recoil spectrum.
We present comparisons to the full toy-MC simulation computation result for several recoil spectra. For heavy WIMPs, the limit computed with the approximate likelihood is typically conservative and within $10\%$ of the full-likelihood computation, while lower-energy recoil spectra see a higher spread around the full-likelihood upper limit of up to $\sim$\,30\%. 
Finally, we extend this work by including the XENONnT 20 tonne-year sensitivity projection~\cite{xenonnt_sensitivity}, with 1000 toy-MC simulation binwise likelihoods, so that the sensitivity of this projection can be evaluated for any NR signature.

In Section~\ref{sec:xenon1t} we give an overview of the XENON1T NR search, highlighting the analysis dimensions used in the inference, and in Section~\ref{sec:migrationmatrix} we discuss the response to NRs. We present our statistical model in Section~\ref{sec:stats_model} and the exact methodology in Section~\ref{sec:profiling}. Section~\ref{sec:inference} details how to use this approach for approximate limits, and provides estimates of the bias and variance of the method for a selection of NR recoil spectra.

\section{XENON1T Nuclear Recoil Search}\label{sec:xenon1t}
\label{sec:dimensions}

XENON1T was designed and optimized to detect the low-energy NRs expected from WIMPs recoiling off xenon nuclei~\cite{xenon1t_instrument}. Its primary detector was a dual phase xenon time projection chamber (TPC) containing 2~tonnes of  instrumented liquid xenon, observing scintillation and ionization charges from interactions in the target. 
Prompt scintillation light is observed from the recombination or de-excitation of xenon ions or dimers, respectively, and is referred to as the S1 signal. 
Ionization electrons are drifted to the liquid-gas interface at the top of the detector by means of a drift field applied between a cathode electrode at the bottom of the chamber and a grounded gate electrode just below the liquid-gas interface. 
The electrons produce scintillation light proportional to the charge, referred to as the S2 signal, when they are extracted into the gas by a higher extraction field. 
Xenon scintillation was observed by 248 photomultiplier tubes\\ (PMTs) arranged in two arrays at the top and bottom of the detector. 
The x-y position of the interaction was inferred from the pattern of S2 photons observed by the top PMT array, while the time separation between the S1 and the S2 signal indicated the z-depth. With access to the full 3D position information we could fiducialize the detector volume, selecting only the innermost 1.3\,tonne xenon volume where contributions from radioactivity in the detector materials are minimized.

\begin{figure}[t]
    \centering
    \includegraphics[width=0.98\columnwidth]{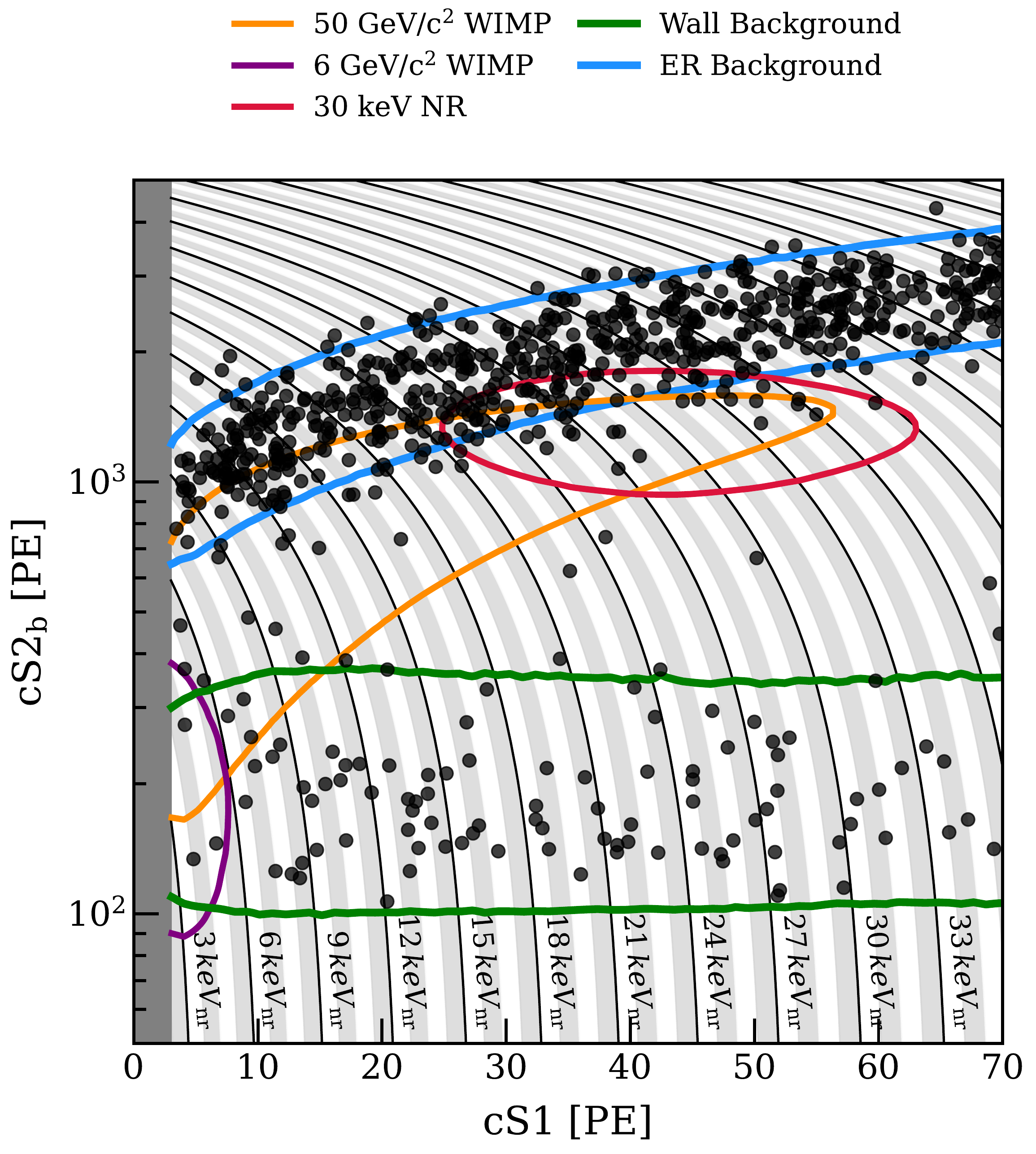}
    \caption{Scatter-plot of the XENON1T NR search dataset in \cSone and \cStwob. Gray lines indicate the 80 bins in reconstructed NR energy. Coloured contours indicate 1$\sigma$ contours for background and signal models: Blue and green contours show the ER and wall background models, the purple and orange contours show the $6~\mathrm{GeV}/\mathrm{c}^2$ and  $50~\mathrm{GeV}/\mathrm{c}^2$ spin-independent WIMP signal models, and the red a $30~\mathrm{keV}$ monoenergetic NR recoil.} 
    \label{fig:scatter_in_erec}
\end{figure}

Both S1 and S2 signals were corrected to account for the detector's position dependent light collection efficiency~\cite{xenon1t_analysis1}, and in the case of the S2 we also corrected for electron attachment to impurities in the liquid xenon volume as the electrons are drifted upwards. These corrected S1 and S2 variables are named \cSone and \cStwo. 

The relative size of the ionisation and scintillation signals, and therefore \cSone and \cStwo, depends on whether the incident particle scattered off the nucleus (NR) or an electron (ER) of a xenon atom. In Figure~\ref{fig:scatter_in_erec} the predicted 1$\sigma$ contour for interactions of a 50 (6) GeV/c$^2$ WIMP, which is expected to interact with the xenon nucleus, producing NRs, is shown in orange (purple). We also illustrate the signal expectation from a mono-energetic 30\,keV NR line in red. The 1$\sigma$ contour of the ER background is shown in blue, demonstrating the separation between nuclear and electronic recoils in XENON1T. Shown in green is the 1$\sigma$ contour of the ``wall" background, which is discussed at the end of this section.

WIMPs are expected to scatter at most once off a target nucleus due to their small interaction  cross sections, therefore XENON1T optimized its search strategy to look for single scatter NR events. The analysis space spans from 3 to 70\,photoelectrons (PE) in the \cSone space, where the lower boundary is driven by the detection efficiency, and the upper boundary is chosen to include the bulk of the expected WIMP signal. We use the light observed in the bottom PMT array to determine magnitude of the  position corrected S2s, referred to as \cStwob, due to the more uniform response of this array in the x-y plane. The \cStwob space is chosen to fully contain the expected background and signal models in our chosen \cSone region and spans from 50 to 7940\,PE, corresponding to approximately 1.5 to 250\,electrons.

\sloppy
The full XENON1T exposure was collected in two science campaigns, SR0 and SR1, between November 2016 and February 2018, with drift fields of 120\,V/cm and 81\,V/cm, respectively. Continual purification of the xenon improved the electron lifetime from 380\,\micros\ at the start of SR0 to\\ $\sim$\,650\,\micros\ at the end of SR1.
The final data, after quality selections detailed in~\cite{xenon1t_analysis1} and fiducialization consisted of 739 events in a $1~\mathrm{tonne}$-year exposure, shown in Figure~\ref{fig:scatter_in_erec} as grey circles.
\fussy

The response of the detector to low-energy ER and NR interactions was calibrated with \radonttz, the decay products of which produce low-energy beta-decays, and $^{241}$AmBe and deuterium-deuterium fusion generator neutron sources. We used a detector response model based on a fast detector simulation to fit the calibration data and model ER and NR sources in XENON1T~\cite{xenon1t_analysis2}.

Background models for five sources of interactions within XENON1T were considered, detailed in~\cite{xenon1t_analysis2}. The largest background is ERs induced by the $^{214}$Pb decay product of $^{222}$Rn or decays of $^{85}$Kr. The second largest background expectation is referred to as the ``wall" background. These are events which occur close to the polytetrafluoroethylene walls of the detector, and consequently lose a portion of the ionization electrons to the wall as they drift upwards. The lower S2 signal, observed close to the detector edge will result in larger position reconstruction errors, and this population will therefore bleed into the fiducial volume. For this reason, we include the radius, denoted by R, as an analysis dimension along with \cSone and \cStwob for the background and signal models. The 1$\sigma$ contours of these two dominant backgrounds is shown in Figure~\ref{fig:scatter_in_erec} in blue (ERs) and green (wall). The other backgrounds considered are radiogenic neutrons from detector materials, coherent elastic neutrino-nucleus scattering (CE$\nu$NS) of solar $^8$B neutrinos, and accidental pairing of lone S1 and S2 signals. 

\section{Analysis variables and Detector Response}
\label{sec:migrationmatrix}
Previous XENON1T searches for WIMP interactions~\cite{xenon1t_analysis1,xenon1t_analysis2} directly used the observed \cSone and \cStwob variables as described in Section~\ref{sec:xenon1t}. Since the total number of quanta produced is dependent on the original energy deposition, the number of prompt scintillation photons and ionization electrons, observed as the \cSone and \cStwob, respectively,are intrinsically anti-correlated. Additionally, the fraction of quanta observed as ionization electrons or scintillation photons is energy dependent. Thus for a given \cSone selection, different NR energies yield different distributions in \cStwob space. To reduce the dependence on the recoil energy, we transform our analysis space to explicitly feature reconstructed energy as one dimension. 

\subsection{Reconstructed Energy}
The reconstructed ER energy ${\Erec}_\mathrm{ER}$ of the original interaction can be obtained from \cSone and \cStwob quantities as:

\begin{align}
    {\Erec}_\mathrm{ER}    (\cSone,\cStwob) \equiv  W \cdot [\cSone/g_1 + \cStwob/g_2],
\end{align}
where $W$=13.7\,eV is the average amount of energy required to produce one electron or photon in xenon~\cite{DahlThesis}.
The detector dependent quantities $g_1$ and $g_2$ represent the number of photoelectrons observed in the PMT arrays per emitted scintillation photon and the number of photoelectrons observed per extracted electron respectively. 

Since the approximate likelihood will be presented in bins of \Erec, it is necessary that other analysis dimensions are as independent of recoil energy as possible. 
Therefore, we also introduce \Erecperp,
\begin{equation}
      \Erecperp(\cSone,\cStwob) \equiv W \cdot [\cSone/g_2 - \cStwob/g_1],
\end{equation}
which is constructed so that ${\Erec}_\mathrm{ER}$ and $\Erecperp$ contours are perpendicular. 

Performing the analysis in \Erec, \Erecperp coordinates rather than in \cSone, \cStwob is only a coordinate transformation, and does not affect the XENON1T inference results. 

In order to obtain the reconstructed recoil energy for NR events (${\Erec}$), one must also account for the energy dependent quenching effect, where NR energy is lost to unobserved heat. We estimate the quenching magnitude at a given energy from an empirical comparison between the true NR energy and the reconstructed ER energy using the detector response model described in~\cite{xenon1t_analysis1}. The constant NR energy lines obtained from the above procedure are shown in Figure~\ref{fig:scatter_in_erec} as gray shaded bands. 

\subsection{Migration Matrix}\label{subsec:migrationmatrix}

\begin{figure}[t]
\includegraphics[width=0.98\columnwidth]{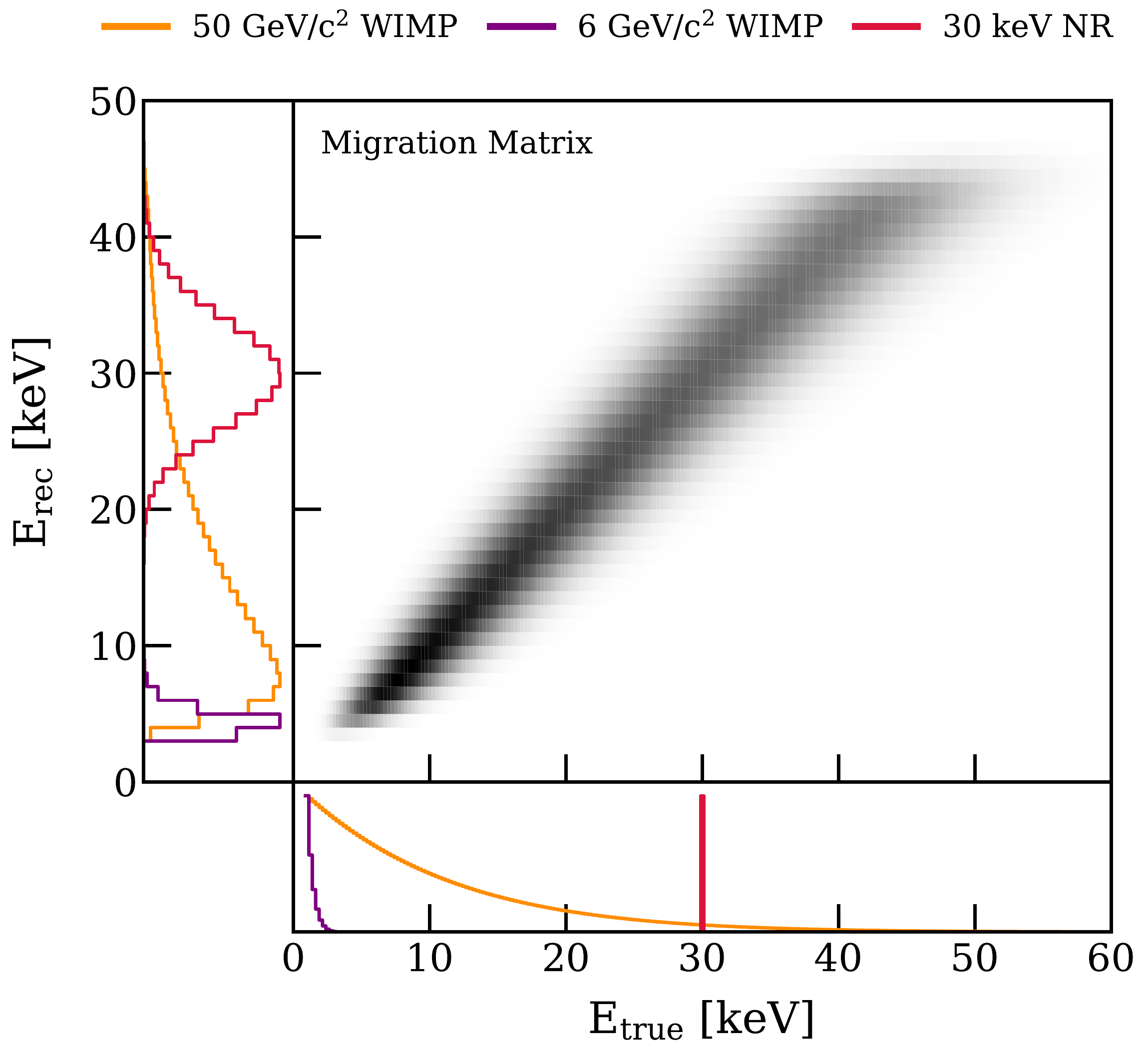}
\caption{\label{fig:migrationmatrix} Illustration of the migration matrix included in the data release~\cite{xenon1t_binwise_data}, as defined in equation~\ref{eq:migrationmatrix}, showing the conversion between true NR recoil energy and the reconstructed energy. The bottom panel shows the true NR spectrum of a $30~\mathrm{keV}$ line in red, and spin-independent (SI) WIMP recoil spectra for a $6~\mathrm{GeV}/\mathrm{c}^2$ and $50~\mathrm{GeV}/\mathrm{c}^2$ WIMP in purple and orange, respectively, all with arbitrary normalisation. The left panel shows the same spectra in reconstructed energy after multiplication with the migration matrix. The matrix is normalized such that selections in \Erec account for our overall detection efficiency}
\end{figure}

In order to convert an arbitrary NR spectrum into the reconstructed energy spectrum expected to be observed in\\ XENON1T we account for detector effects. The complete detector response model, derived from fits to calibration data and accounting for detection efficiency, resolution and correction effects is described in~\cite{xenon1t_analysis1}. Using this model, we calculate the spread in reconstructed energy space of a fine grid of true NR recoil energies. The migration matrix is shown in Figure~\ref{fig:migrationmatrix}, where the components of the migration matrix 

\begin{equation}
    \migmat_{r,t} = P(\Erec~\text{in bin } r\mid \Etrue~\text{in bin\ } t)\label{eq:migrationmatrix},
\end{equation}
represent the probability for a NR recoil in some true recoil bin to be reconstructed in a given reconstructed energy bin.
The transformation of the true recoil energy spectra for a 6 and 50\,GeV/c$^2$ WIMP into bins in reconstructed energy space is shown in purple and orange respectively. Also shown is the transformation of a mono-energetic 30\,keV line, illustrating the broadening of the signal spectrum from detector effects.

\section{Statistical model}
\label{sec:stats_model}
We use a profiled log-likelihood ratio test statistic and toy-MC simulations of the test statistic distribution to compute discovery significances and confidence intervals.
The likelihood $\L_\mathrm{total}$ used for NR searches with XENON1T is presented in~\cite{xenon1t_analysis2}. It is a product of:
\begin{itemize}
    \item $\L_\mathrm{SR}^\mathrm{sci}(s,\vec{\theta}\given\vec{x})$: unbinned, extended likelihood terms in three analysis dimensions: \cSone, \cStwob and \radius,~for the two science data-taking periods, labeled SR0 and SR1 (indexed with SR). The likelihood is a function of the signal strength parameter $s$, and the set of nuisance parameters $\vec{\theta}$, and is evaluated for the data $\vec{x}$.
    \item $\L_\mathrm{SR}^\mathrm{cal}(\vec{\theta}\given\vec{x})$: unbinned, extended likelihood terms in two analysis dimensions; \cSone and \cStwob for the $^{220}\mathrm{Rn}$ calibration data taken for each science data-taking period. Since this calibration source is uniformly distributed in the detector, \radius~is not included.
    \item $\L^\mathrm{anc}(\vec{\theta}\given\vec{x}_\mathrm{anc})$: terms representing ancillary measurements of background rates and the signal detection efficiency, with $\vec{x}_\mathrm{anc}$ being the ancillary measurements.
\end{itemize} 

%\subsection{Likelihood}
The aim of this paper is to present an approximate likelihood applicable to any NR signal in an easily publishable format. 
To that end, we first reparameterise the signal and background models to be in \Erec, \Erecperp~and \radius, and write separate likelihood terms, primed to mark the reparameterisation, $\L^\mathrm{sci\prime}_\mathrm{r,SR}$ for bins $r$ in reconstructed energy. 
These two changes leave the likelihood unaltered (up to a constant factor)
\begin{align}
\L^\mathrm{tot\prime}(s,\vec{\theta})  =&  \prod_r \prod_\mathrm{SR} \L^\mathrm{sci\prime}_\mathrm{r,SR}(s,\vec{\theta}) \nonumber\\
 \times&\L_\mathrm{SR}^\mathrm{cal\prime}(\vec{\theta})\times        \L^\mathrm{anc}(\vec{\theta}).\label{eq:llsr0binned}
\end{align}
The per-bin science data likelihood for bin $r$ with observed events $N_r$ and lower and upper edges $\Erec{}_{,d}$ and $\Erec{}_{,u}$ is

\begin{align}
    \L^\mathrm{sci\prime}_\mathrm{r,SR}(s,\vec{\theta}) =&Poisson(N_r\given\mu_r^\mathrm{tot}(s,\vec{\theta}))\nonumber\\
    \times&\prod_{i\in S_r} f^\mathrm{tot\prime}(\Erec{}_{,i}, \Erecperp{}_{,i}, \radius_i \given s,\vec{\theta})\label{eq:lsciprimeb}
\end{align} 
where $f^\mathrm{tot\prime}(\Erec,\Erecperp,\radius\given s,\vec{\theta})$ is the total probability density function (PDF) in the transformed analysis variables, and $S_r \equiv \{i\given \Erec{}_{,d}<\Erec{}_{,i}<\Erec{}_{,u}\}$ is the set of events in the science run with \Erec in bin $r$. The total expected number of events in each bin $r$, and the expectation from each source $j$ in that bin are defined as

\begin{align}
    \mu_r^\mathrm{tot}(s,\vec{\theta}) \equiv& \sum_j \mu_{j,r}(s,\vec{\theta})\\
    \mu_{j,r}(s,\vec{\theta}) \equiv& \mu_j(s,\vec{\theta}) \nonumber\\ 
    & \times\int_{\Erec{}_{,d}}^{\Erec{}_{,u}}
    \left(\int f_j^\prime (\Erec, \Erecperp,\radius\given s,\vec{\theta}) \mathrm{d}\Erecperp \mathrm{d}\radius\right)\mathrm{d} \Erec,
\end{align}

where $\mu_j(s,\vec{\theta})$ and $f_j^\prime(\Erec, \Erecperp,\radius\given s,\vec{\theta})$ are the expected number of events and the total PDF of source $j$, respectively.

The first approximation we make is to replace the PDF in each bin by the averaged PDF in that bin, we will denote this change with double primes,
\begin{align}
    f_{j,r}^{\prime\prime}(\Erecperp,R\given s,\vec{\theta}) \equiv \frac{\mu_{j,r}(s,\vec{\theta})}{\mu_j(s,\vec{\theta})}\nonumber \\
    \times \int_{\Erec{}_{,d}}^{\Erec{}_{,u}} f_j^\prime(\Erec, \Erecperp,R\given s,\vec{\theta}) \mathrm{d} \Erec \label{eq:fbj}
\end{align} 
and the science likelihood for the bin to one using this averaged PDF,
\begin{align}
    \L^\mathrm{sci\prime\prime}_\mathrm{r,SR}(s,\vec{\theta}) =&Poisson(N_r\given \mu_r^\mathrm{tot}(s,\vec{\theta}))\\
    \times&\prod_{i\in S_r} f^\mathrm{tot\prime\prime}(\Erecperp{}_{,i}, \radius_i \given s,\vec{\theta})\nonumber\\
     \L^\mathrm{sci\prime\prime}_\mathrm{r}(s,\vec{\theta}) \equiv& \prod_\mathrm{SR}\L^\mathrm{sci\prime\prime}_\mathrm{r,SR}(s,\vec{\theta}).
\end{align} 
The total approximate likelihood is the product of each binwise contribution times the calibration and ancillary constraint terms,
\begin{align}
\L^\mathrm{tot\prime\prime}(s,\vec{\theta})  =&  \prod_r \left( \L^\mathrm{sci\prime\prime}_\mathrm{r}(s,\vec{\theta})\right)\nonumber\\
\times&\L^\mathrm{cal\prime}(\vec{\theta})\times\L^\mathrm{anc}(\vec{\theta}).
\end{align}
\section{Binwise profiling}\label{sec:profiling}
For the binwise-averaged likelihoods to be a good approximation to the unbinned likelihood, the bins must be small with respect to the XENON1T resolution. In Section~\ref{subsec:binbias}, we choose the bin number $n$ to minimise bias and maximise accuracy. To produce a likelihood for any signal shape, we wish to compute profiled likelihood ratios for each bin separately. However, the chosen binning is so narrow that many nuisance parameters in $\vec{\theta}$, for instance the normalisation of the wall background, cannot be constrained in each bin separately. In practice, no nuisance parameter is strongly pulled from its best-fit value in the original XENON1T upper limit computation. Therefore, our second approximation is to first compute $\hat{\vec{\theta}}_0$, the value of the nuisance parameters that optimises $\L^\mathrm{tot\prime\prime}(0,\vec{\theta})$, and fix the nuisance parameters to this value. The exception is the ER mismodelling term (and therefore also the ER normalisation) that requires special attention.

Over- or under-estimating a signal-like tail of the background model would bias results towards too-strict limits or spurious discoveries, respectively. Therefore, the XENON1T WIMP search likelihood~\cite{xenon1t_analysis2} includes an ER mismodelling term~\cite{safeguard} that takes the form of a signal-like component added to the ER model,
\begin{align}
    f_\mathrm{ER}(x) \rightarrow \gamma(\alpha)\times\mathrm{max}\left[(1-\alpha) f_\mathrm{ER}(x) + \alpha f_\mathrm{SIG}(x),0\right]
\end{align}
where $f_\mathrm{ER}(x)$, $f_\mathrm{SIG}(x)$ are the PDFs in $x$ of the ER background and (WIMP) signal, respectively, $\alpha$ is the size of the ER mismodelling term and $\gamma(\alpha)$ is a normalisation term to ensure that the total PDF is normalized even for negative $\alpha$.
Since this term depends on the signal model considered, it cannot be determined by the background-only fit, and must be profiled per bin. The total ER distribution used in the likelihood becomes 
\begin{align}
    &f_\mathrm{ER,r}(\vec{x} \given \alpha_r) \equiv\nonumber \\
    &\begin{cases}
    %\begin{aligned}
    \gamma(\alpha_r)\times\mathrm{max}[0,((1-\alpha_e)f_\mathrm{ER}(x\given\hat{\vec{\theta_0}}) + \\
    \alpha_r \times f_\mathrm{SIG}(x))
    ],  \ \ \ \ \ \ \ \ \ \ \ \ \ \ \ \  \text{if } \Erec{}_{,d}<\Erec\leq\Erec{}_{,u}\\
    %\end{aligned}\\
    \gamma(\alpha_r)\times f_\mathrm{ER}(\vec{x}\given\hat{\vec{\theta_0}}),            \ \ \ \ \ \ \ \text{otherwise,}
\end{cases}\label{eq:er_with_mm_in_bin}
\end{align}
which is used both in the calibration and science data likelihoods. Each bin has its own mismodelling component, parameterized with $\alpha_r$, which therefore can more freely fit the calibration data shape, resulting in an improved fit. The total calibration PDF is the sum of $f_\mathrm{ER}(\vec{x}\given\hat{\vec{\theta}}_0)$ and the accidental background component making $\L^\mathrm{\prime\prime\mathrm{cal}}(\alpha_r)$ have the same form as equation~\ref{eq:lsciprimeb}. Since the mis-modelling term only affects the shape of the background, the normalisation in the calibration term is fixed to the best-fit value.

Using the ER model of equation~\ref{eq:er_with_mm_in_bin} and the best-fit nuisance parameters for the no-signal fit $\hat{\vec{\theta}}_0$, thereby fixing $\L^\mathrm{anc}$, we construct the likelihood in each bin of reconstructed energy,

\begin{align}
\L^\mathrm{tot\prime\prime}(s,\alpha_r,\mu^\mathrm{ER}_r)_r  =
\L^\mathrm{sci\prime\prime}(s_r,\alpha_r,\mu^\mathrm{ER}_r,\hat{\vec{\theta}}_0)\times\L^\mathrm{cal\prime\prime}(\alpha_r,\hat{\vec{\theta}}_0).
\end{align}
Here, $s_r$ is the signal expectation in each reconstructed energy bin $r$, which relates to the expectation in each bin of true energy $t$ via the migration matrix
\begin{align}
    s_r = \sum_t \migmat_{r,t} \cdot s_t,
\end{align}
and the signal expectation in each true energy bin $t$ in turn is given by
\begin{align}
    s_t = s \int_{\Etrue{}_{,d}}^{\Etrue{}_{,u}} g(\Etrue) \mathrm{d} \Etrue,
\end{align}
where $g(E)$ is the signal PDF in true recoil energy  $\Etrue$, and $s$ the expected number of true signal events.

\begin{figure}[t]
\includegraphics[width=0.98\columnwidth]{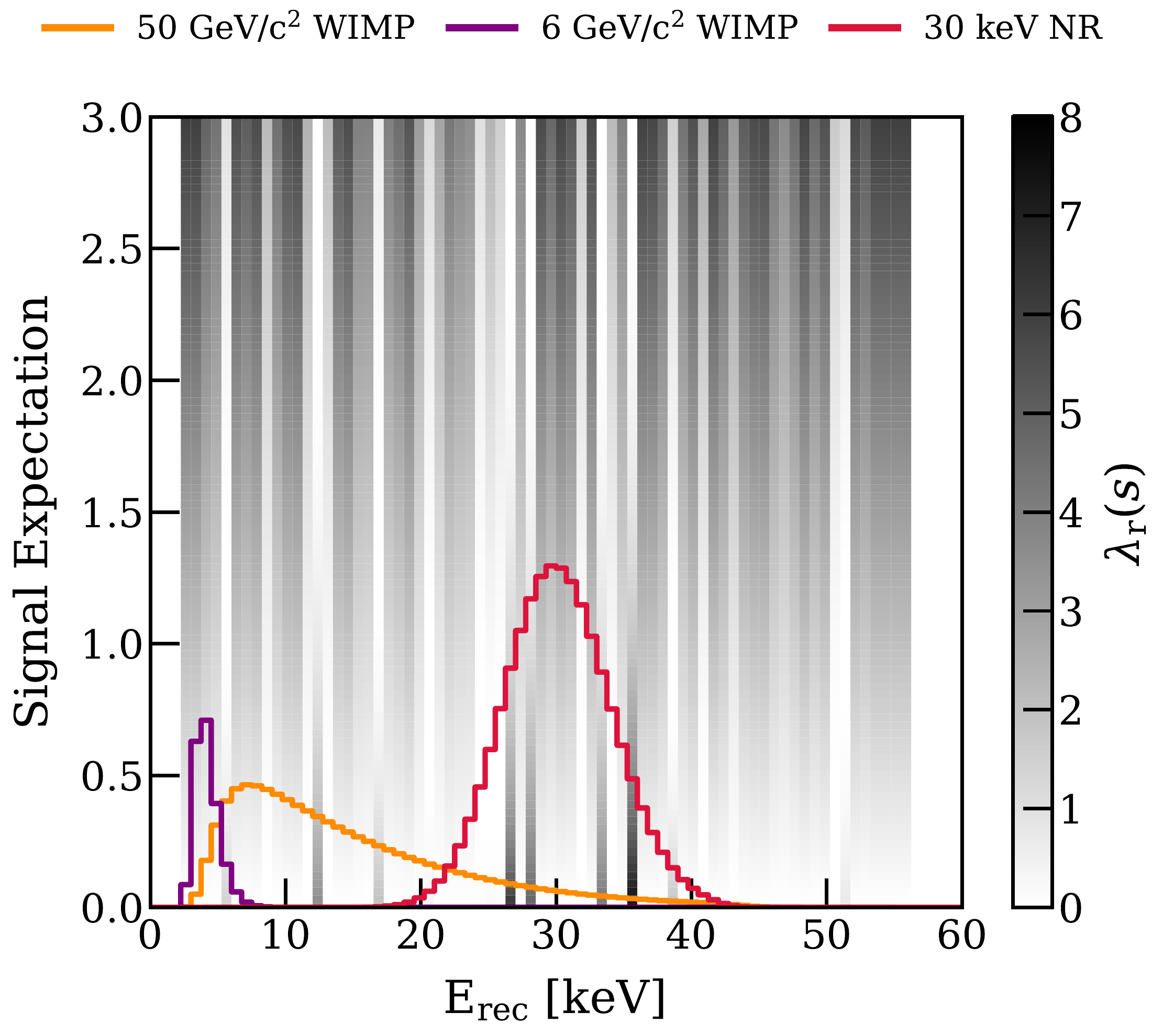}
\caption{\label{fig:binnedll} Illustration of the binwise, profiled log-likelihood $\lambda_\mathrm{r}(s)$ for bins in reconstructed NR energy. The total approximate likelihood is obtained by summing over the entry in each reconstructed energy bin at the expected signal, as in equation~\ref{eq:approximatell}. The purple, orange and red lines indicate the expectation values in each bin for a $6~\mathrm{GeV}/\mathrm{c}^2$ and a $50~\mathrm{GeV}/\mathrm{c}^2$ spin-independent WIMP signals  and a $30$\,keV NR line signal respectively, at their respective upper limits derived from the XENON1T dataset.  White bins at the highest and lowest reconstruction energies reflect bins for which the migration matrix is 0.}
\end{figure}

The binwise profiling follows the approach in~\cite{Ackermann:2015zua} and~\cite{xenon100_eft}, where the likelihood is profiled separately in sections of the analysis variable space. The profiled likelihood in each bin is
\begin{align}
    \lambda_\mathrm{r}(s) = -2\times\log\left(\frac{\L^\mathrm{tot\prime\prime}(s,\hat{\hat{\alpha_r}},\hat{\hat{\mu}}^\mathrm{ER}_r)}{\L^\mathrm{tot\prime\prime}(\hat{s},\hat{\alpha_r},\hat{\mu}^\mathrm{ER}_r)}\right)\label{eq:binprofilellr}
\end{align}
where $\hat{s},\hat{\alpha_r},\hat{\mu}^\mathrm{ER}_r$ is the signal expectation value, mismodelling fraction and ER rate that maximises the likelihood, and $\hat{\hat{\alpha_r}},\hat{\hat{\mu}}^\mathrm{ER}_r$ maximise the conditional likelihood. Figure~\ref{fig:binnedll} shows the profiled binwise likelihood for each bin as function of $s$. The per-bin likelihoods show the expected fluctuation from a lower-statistics sample-- some prefer a positive signal, others no signal. To compute a full result, they must be combined into one likelihood. 

\section{Inference using the binwise likelihood}
\label{sec:inference}
Using the energy migration matrix defined in Section~\ref{sec:migrationmatrix} to compute bin-wise signal expectations $s_r$, together with the likelihood ratio for each bin defined in equation~\ref{eq:binprofilellr}, we can write our approximation of the log-likelihood written in equation~\ref{eq:llsr0binned} 
\begin{align}
    \Lambda_\mathrm{tot}(s) = \sum_r{\lambda_\mathrm{r}(s_\mathrm{r})}\label{eq:approximatell}
\end{align}
and the corresponding log-likelihood ratio 
\begin{align}
    \lambda_\mathrm{tot}(s) = \Lambda_\mathrm{tot}(s) - \Lambda_\mathrm{tot}(\hat{s})\label{eq:approximatellr}.
\end{align}
Using this approximate likelihood induces only a moderate systematic and random error in confidence intervals with respect to the ones computed with the full, computationally much slower XENON1T likelihood. Best-fit and upper limits are then computed using the standard asymptotic formulae~\cite{wilks,CowanAsymptotic}.

\setlength{\tabcolsep}{4pt} 
\bgroup
\def\arraystretch{1.3}%
\begin{table}[t]
    \centering
    \begin{tabular}{r|ccc}
   &  $n=60$ Bins &  $n=80$ Bins &  $n=120$ Bins\\
   \hline
\multicolumn{1}{l|}{Flat Spectrum}  & ${ 1.01 } ^ {+ 0.08 } _{-  0.06 }$ & ${ 1.01 } ^ {+ 0.07 } _{-  0.05 }$ & ${ 1.01 } ^ {+ 0.07 } _{-  0.04 }$\\
\multicolumn{1}{l|}{NR Lines}\\
$3~\mathrm{keV}$   & ${ 1.06 } ^ {+ 0.33 } _{-  0.16 }$ & ${ 1.04 } ^ {+ 0.29 } _{-  0.15 }$ & ${ 1.03 } ^ {+ 0.23 } _{-  0.15 }$\\
$5~\mathrm{keV}$   & ${ 1.13 } ^ {+ 0.24 } _{-  0.14 }$ & ${ 1.11 } ^ {+ 0.19 } _{-  0.13 }$ & ${ 1.11 } ^ {+ 0.17 } _{-  0.12 }$\\
$7~\mathrm{keV}$   & ${ 1.14 } ^ {+ 0.19 } _{-  0.14 }$ & ${ 1.13 } ^ {+ 0.16 } _{-  0.12 }$ & ${ 1.13 } ^ {+ 0.15 } _{-  0.12 }$\\
$10~\mathrm{keV}$   & ${ 1.11 } ^ {+ 0.15 } _{-  0.12 }$ & ${ 1.11 } ^ {+ 0.14 } _{-  0.11 }$ & ${ 1.11 } ^ {+ 0.12 } _{-  0.10 }$\\
$20~\mathrm{keV}$   & ${ 1.05 } ^ {+ 0.14 } _{-  0.09 }$ & ${ 1.04 } ^ {+ 0.12 } _{-  0.08 }$ & ${ 1.05 } ^ {+ 0.12 } _{-  0.08 }$\\
$30~\mathrm{keV}$   & ${ 1.04 } ^ {+ 0.11 } _{-  0.09 }$ & ${ 1.03 } ^ {+ 0.10 } _{-  0.08 }$ & ${ 1.04 } ^ {+ 0.10 } _{-  0.08 }$\\
\multicolumn{1}{l|}{SI WIMP signals}\\
$6~\mathrm{GeV}/c^2$ & ${ 1.07 } ^ {+ 0.40 } _{-  0.21 }$ & ${ 1.02 } ^ {+ 0.31 } _{-  0.19 }$ & ${ 1.01 } ^ {+ 0.25 } _{-  0.17 }$\\
$10~\mathrm{GeV}/c^2$ & ${ 1.08 } ^ {+ 0.24 } _{-  0.14 }$ & ${ 1.06 } ^ {+ 0.19 } _{-  0.14 }$ & ${ 1.05 } ^ {+ 0.17 } _{-  0.13 }$\\
$50~\mathrm{GeV}/c^2$ & ${ 1.07 } ^ {+ 0.10 } _{-  0.08 }$ & ${ 1.06 } ^ {+ 0.09 } _{-  0.07 }$ & ${ 1.07 } ^ {+ 0.07 } _{-  0.07 }$\\
$100~\mathrm{GeV}/c^2$ & ${ 1.06 } ^ {+ 0.10 } _{-  0.08 }$ & ${ 1.06 } ^ {+ 0.09 } _{-  0.06 }$ & ${ 1.06 } ^ {+ 0.07 } _{-  0.06 }$\\
    \end{tabular}
     \caption{Table of bias and spread, defined as the median and 1-sigma spread of the ratio between binwise and full likelihood upper limits using 1000 toy-MC simulations.}
    \label{tab:xe1t_validation}
\end{table} 

\subsection{Fidelity of the approximate likelihood method~\label{subsec:binbias}}

Differences between the unbinned and approximate binwise results are the binning in $\Erec$, the per-bin ER mismodelling and profiling, and the slight change in the signal distribution in $\Erecperp$ in individual bins for different signal shapes.

We validated the performance of the binwise likelihood approach by computing upper limits for a range of signal spectra and different numbers of bins in \Erec. 
Table~\ref{tab:xe1t_validation} shows the median ratio between the limits computed with the  approximate and full likelihood, and errors corresponding to the $15$th and $85$th percentiles of the ratio between the two.
Increasing the number of bins beyond 80 bins between $0$ and $60~\mathrm{keV}$ did not markedly improve either the bias or spread of the upper limits for the binwise likelihood. Therefore we choose to report the result of this work using 80 bins in reconstructed energy space.
For heavy WIMPs, the bias and errors are both on the order of $10\%$. The more peaked low-mass WIMP signals or lower-energy monoenergetic lines, both concentrated in only a few bins, have a larger range of deviation from the full result, up to $30\%$ scatter with respect to upper limits with the full likelihood. 
The bias and errors in Table~\ref{tab:xe1t_validation} give an indication of how well the approximate likelihood should be expected to perform for different signal shapes and energy ranges. 
    
\subsection{Correcting for non-asymptoticity}
\label{subsec:nonasymptotic}
The XENON1T results were computed from test statistic distributions estimated using toy-MC simulations of datasets. This was necessary due to the non-asymptotic nature of the distributions for the low signal-numbers considered~\cite{xenon1t_analysis2}. 
\\Since generating datasets depends on the signal model, this approach must be amended if a similar correction should be applied to the likelihood ratio of equation~\ref{eq:approximatell}. 

\begin{figure}[t]
\includegraphics[width=0.98\columnwidth]{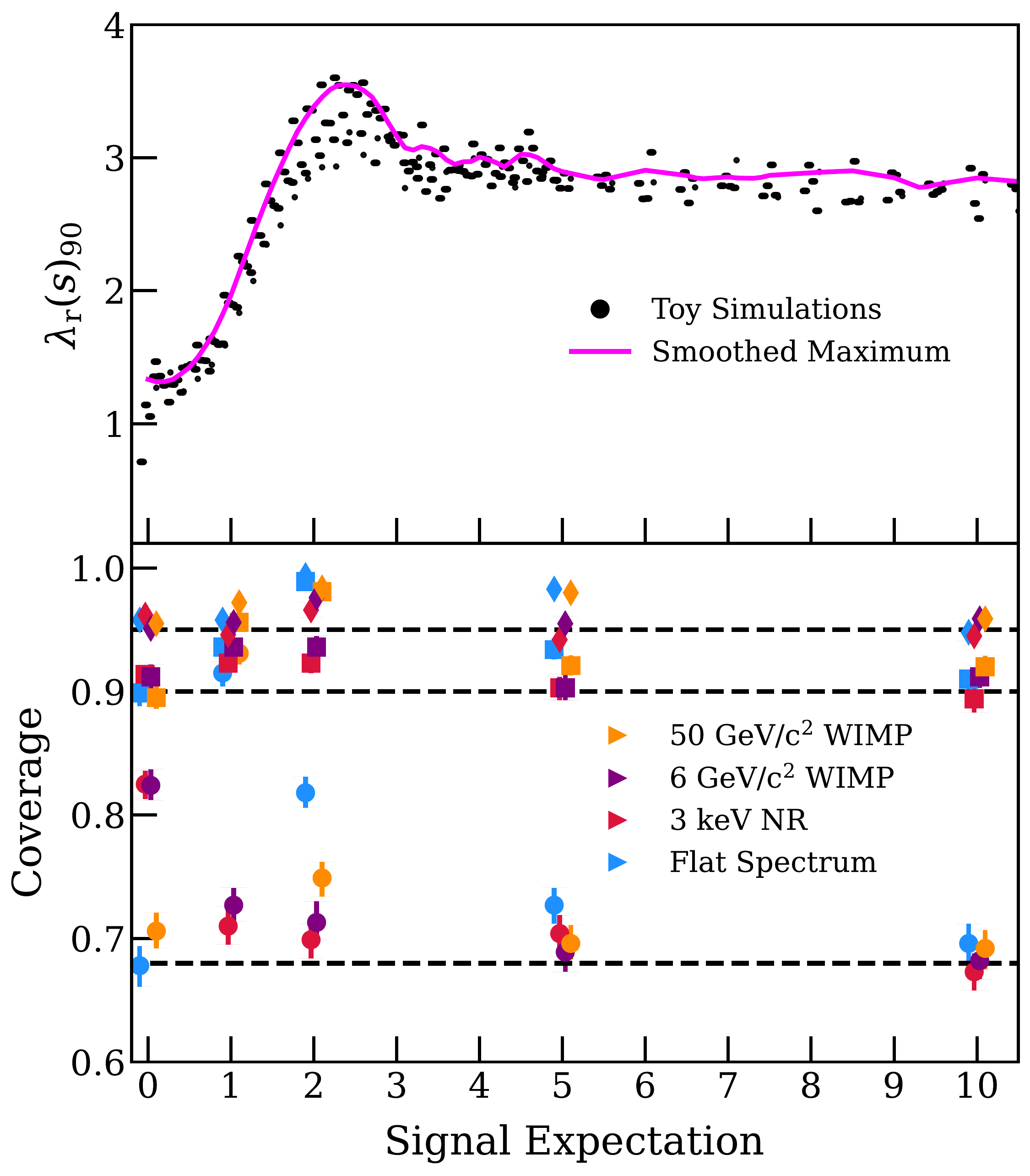}
\caption{\label{fig:nonasymptotic} \textbf{Top}: 
The $90$-percentile threshold of the approximate log-likelihood ratio test statistic as function of the true signal expectation.
Thresholds estimated with toy-MC simulations for a range of monoenergetic signals are shown with black dots, and the magenta line shows the smoothed maximum. 
The threshold converges to the asymptotic value for around $\sim4$ expected signal events.%\\ 
\textbf{Bottom}: The coverage of $95$, $90$ and $68$-percent confidence level upper limits are shown with diamonds, squares and circles, respectively, for five NR recoil spectra: 
Flat (blue), a 3\,keV monoenergetic line (red), a    
$6~\mathrm{GeV}/ \mathrm{c}^2$ SI WIMP (purple) and a 
$50~\mathrm{GeV}/ \mathrm{c}^2$ SI WIMP (orange).
}
\end{figure}

\begin{figure*}[t]
\centering
\subfloat[Spin-independent WIMP-nucleon]{
    \includegraphics[width=0.98\columnwidth]{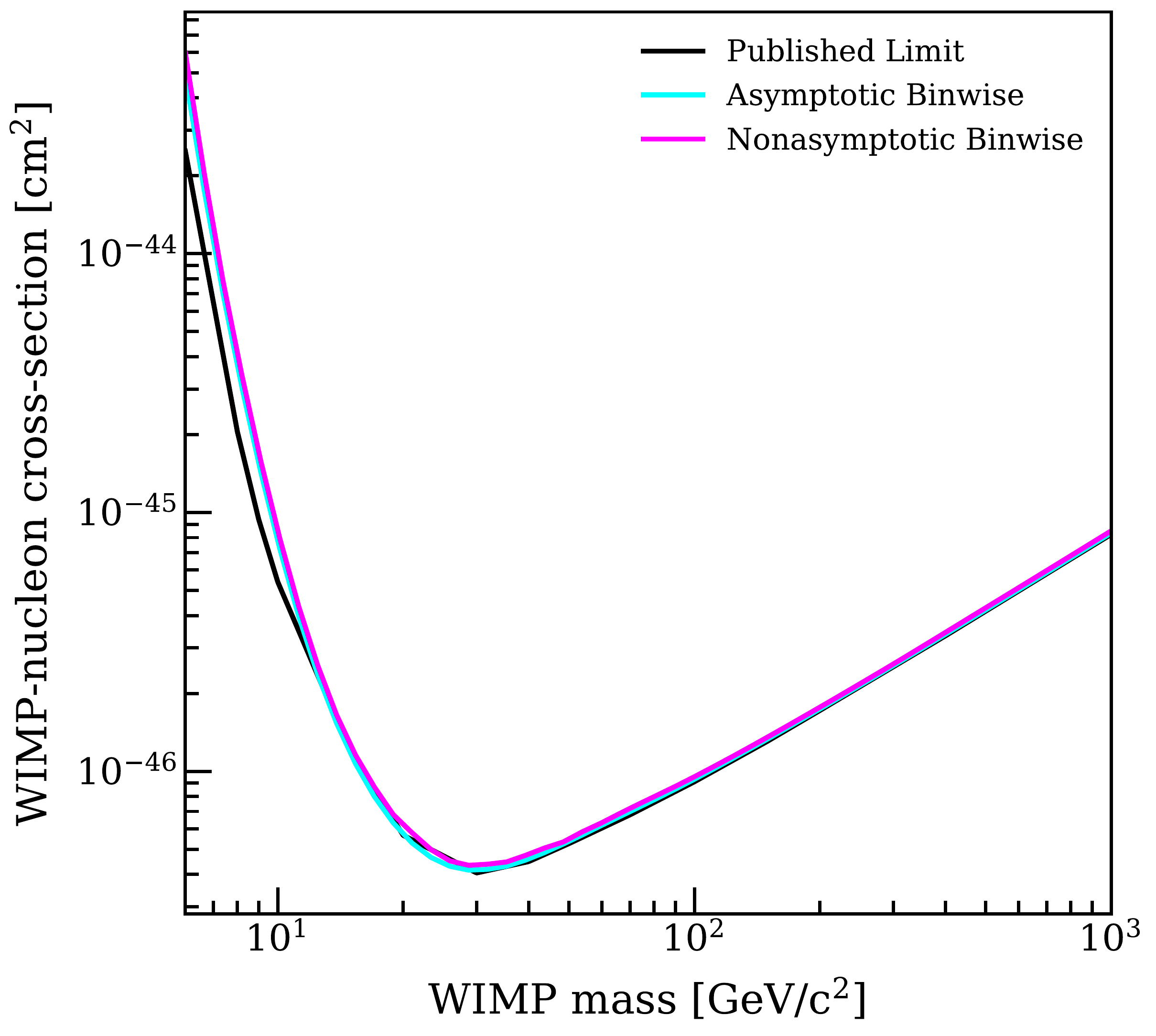}}
\subfloat[Spin-dependent WIMP-proton]{
    \includegraphics[width=0.98\columnwidth]{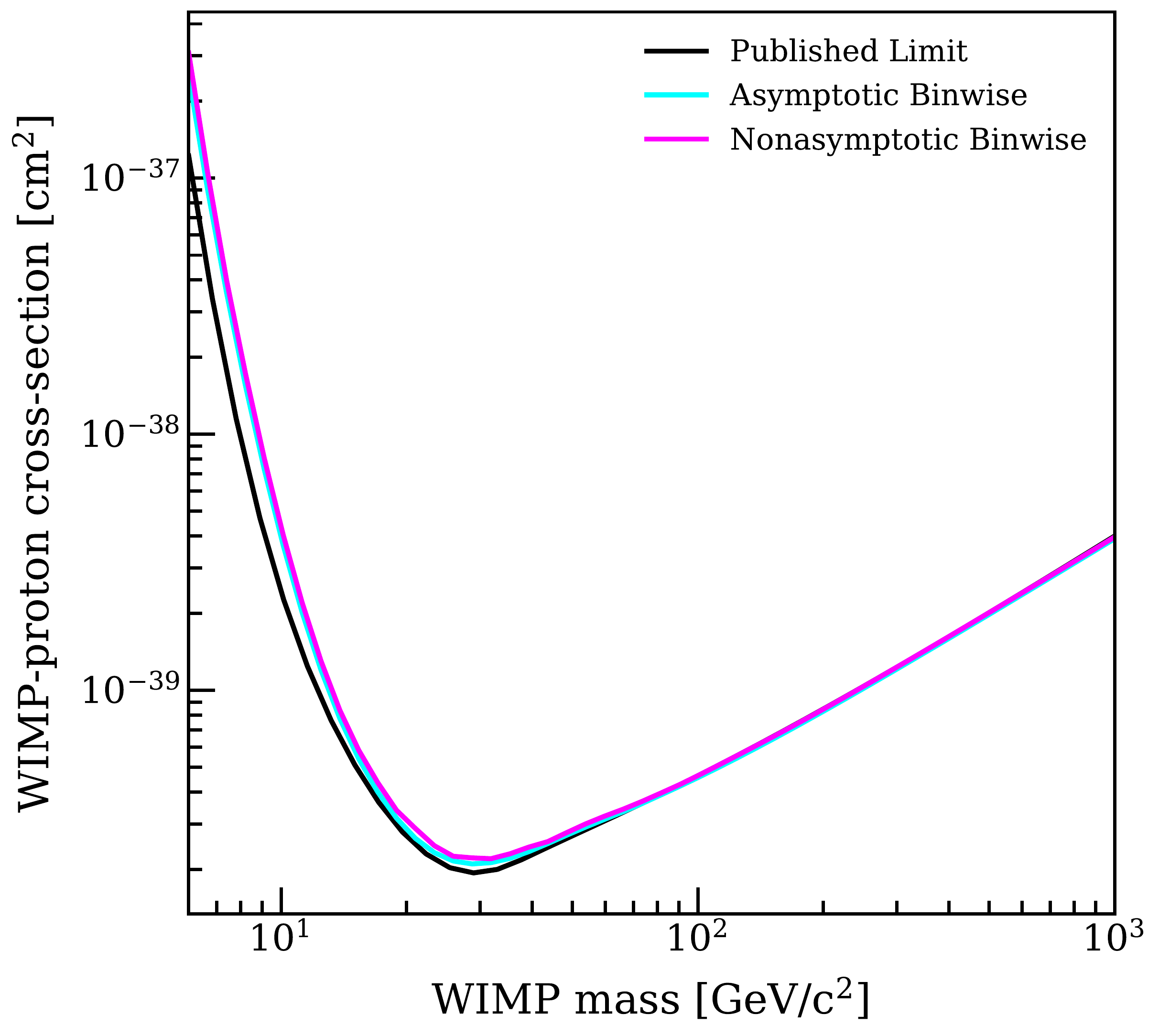}}\\
\subfloat[Spin-dependent WIMP-neutron]{
    \includegraphics[width=0.98\columnwidth]{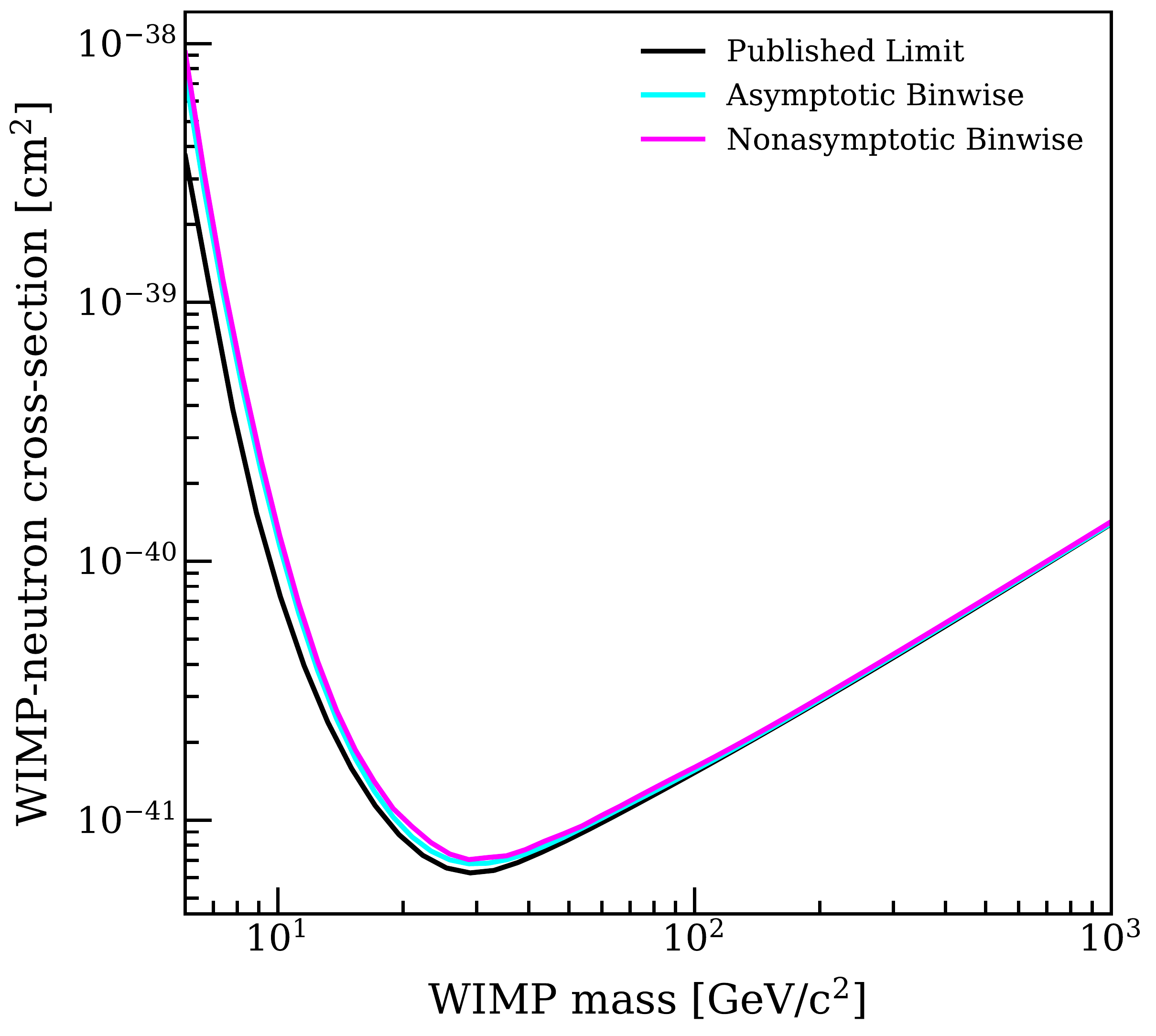}}
\subfloat[WIMP-pion]{
    \includegraphics[width=0.98\columnwidth]{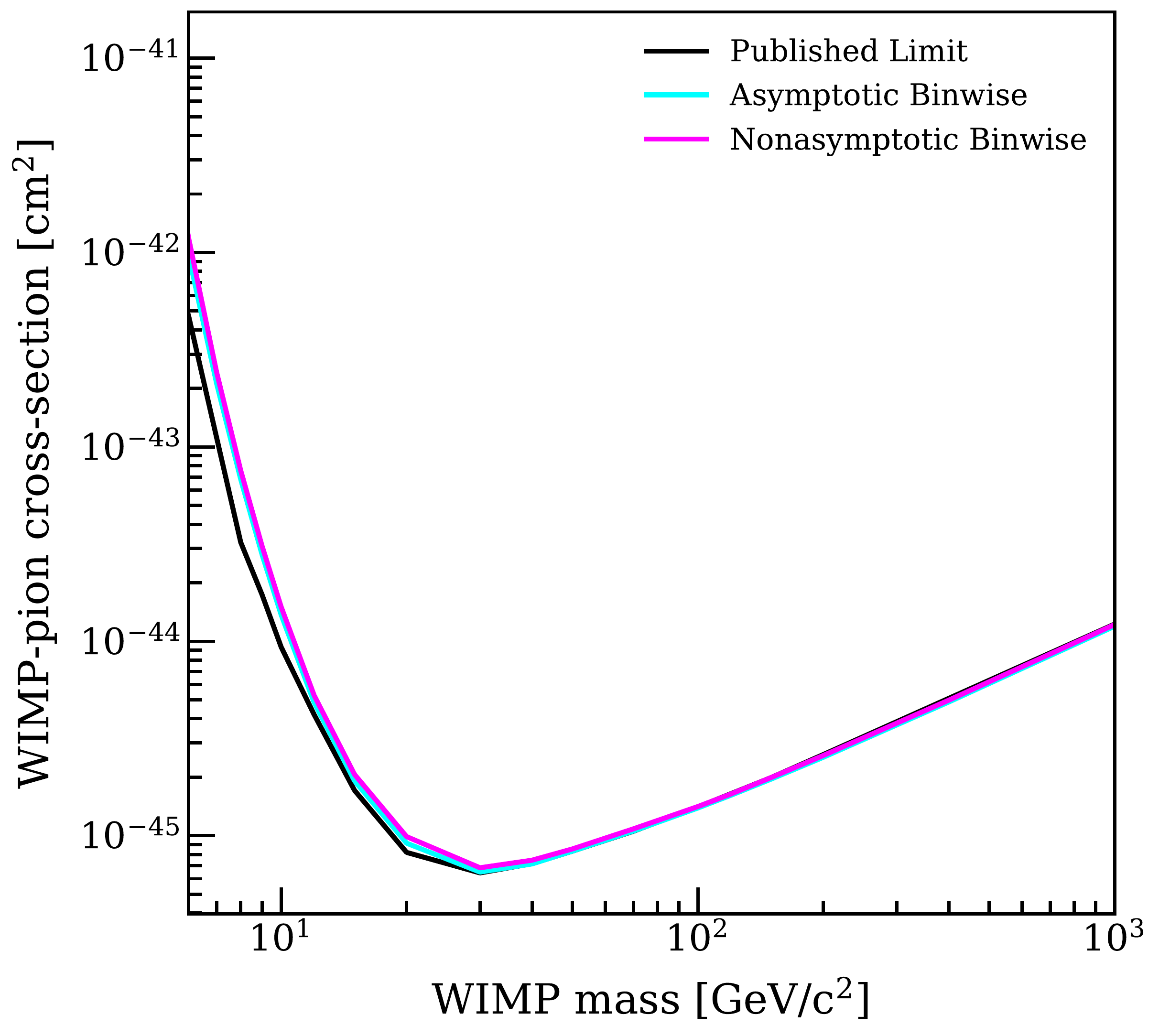}}
\caption{\label{fig:uls}
Comparison between $90\%$ confidence level upper limits from published XENON1T NR searches~\cite{xenon1t_sr1,xenon1t_SD,xenon1t_pion} (black), and limits using the approximate likelihood presented in this work. 
Cyan lines are computed assuming an asymptotic distribution of the test statistic, while magenta lines show the upper limit using the non-asymptotic threshold described in Section~\ref{subsec:nonasymptotic}. 
As in the toyMC studies, the binwise result on data is a good approximation of the full computation for WIMPs with masses $\gtrsim 50~\mathrm{GeV}/c^2$, and gives a conservative result for lower-mass WIMP signals.
}
\end{figure*}

Our approach is motivated by the observation that the non-asymptotic behaviour of the XENON1T likelihood is driven by the signal-to-background discrimination that leaves the signal region almost background free.
Computing the test statistic distribution for a range of monoenergetic NRs will include the best signal-to-background discrimination. Other NR signals will be broader and therefore feature less extreme ER-NR discrimination than these monoenergetic signals. Therefore, we compute the $90$th percentile thresholds of the test statistic for a fine grid of monoenergetic NR signals, and choose the $90$th percentile of all thresholds, to avoid statistical fluctuations, before smoothing this threshold using a Gaussian filter.
Figure~\ref{fig:nonasymptotic} (top) shows the thresholds computed as a function of the signal, while Figure~\ref{fig:nonasymptotic} (bottom) shows the coverage for several signal spectra using the smoothed upper envelope of these thresholds together with equation~\ref{eq:approximatellr} to compute upper limits for several recoil spectra. All show either the nominal coverage, or conservatively over-cover at low signal expectations. We therefore recommend using this threshold rather than the asymptotic $\chi^2$ threshold to compute frequentist confidence intervals. 
In the data release, we include $68$, $90$ and $95$-percentile thresholds.

Upper limits computed with the binwise approximation and with the binwise approximation plus the non-asymptotic threshold are compared with all XENON1T high-mass NR searches in Figure~\ref{fig:uls}. Close agreement is seen except at low masses, where the binwise approximation yields a higher, and thus more conservative, upper limit.  

\section{Summary}
This paper and the accompanying code and data release provide a fast and flexible method to compute approximate results of the XENON1T NR search~\cite{xenon1t_sr1} for any NR spectrum. As many spectra can be tested, care should be taken when interpreting the likelihood to compute discovery significances. On the other hand, we have validated with toy-MC simulations and comparisons with the XENON1T full likelihood that good agreement is found for confidence intervals. We also provide a method to ensure that these confidence intervals have, on average, correct or over-coverage only. In the appendix, we also show how this method can be employed to provide recasts of sensitivity projections, in this case of the $20$ tonne-year XENONnT projection presented in~\cite{xenonnt_sensitivity}.
Together with the XENON1T ER spectral search~\cite{xenon1t_lower,xenon1t_lower_data} and ionisation-only~\cite{xenon1t_s2only,xenon1t_s2only_data} publications, the approximate NR likelihood provides a range of recastable legacy results of the XENON1T experiment.

\section*{Acknowledgments}

We gratefully acknowledge support from the National Science Foundation, Swiss National Science Foundation, German Ministry for Education and Research, \\Max Planck Gesellschaft, Deutsche Forschungsgemeinschaft, Helmholtz Association, Dutch Research Council (NWO), Weizmann Institute of Science, Israeli Science Foundation, Fundacao para a Ciencia e a Tecnologia, R\'egion des Pays de la Loire, Knut and Alice Wallenberg Foundation, Kavli Foundation, JSPS Kakenhi in Japan and Istituto Nazionale di Fisica Nucleare. This project has received funding/support from the European Union’s Horizon 2020 research and innovation programme under the Marie Sk\l odowska-Curie grant agreement No 860881-HIDDeN. Data processing is performed using infrastructures from the Open Science Grid, the European Grid Initiative and the Dutch national e-infrastructure with the support of SURF Cooperative. We are grateful to Laboratori Nazionali del Gran Sasso for hosting and supporting the XENON project.

\appendix
\renewcommand{\thesubsection}{\Alph{subsection}}

\section{XENONnT projection}

The successor to XENON1T, XENONnT, operates under the same principle but is designed with three times the active volume of XENON1T and a lower background~\cite{xenonnt_sensitivity}. As an example of how the approximate likelihood approach can be applied to projections as well, we computed toy-MC binwise likelihoods for 1000 no-signal simulations of the experiment, and compare to the published projections using the full likelihood. Using these, the limit-setting potential of the assumed detector model and exposure can be estimated for any NR signal.

In the projections of its sensitivity~\cite{xenonnt_sensitivity} in a 20 tonne-year exposure, we assumed an electron lifetime of 1\,ms at a drift field of 200\,V/cm. The overall ER background is assumed to be reduced by a factor of 6 from that reported in~\cite{xenon1t_sr1} through selective choice of detector materials and the introduction of a radon distillation column to further reduce the $^{214}$Pb background. A neutron veto is added around the cryostat which contains the time projection chamber of XENONnT in order to suppress the NR background by rejecting 87\% of single-scatter neutron interactions in the active volume. 

\bgroup
\def\arraystretch{1.3}%
\begin{table}[t]
    \centering
    \begin{tabular}{r|c}
   &    80 Bins \\
   \hline
\multicolumn{1}{l|}{Flat Spectrum}& ${ 0.98 } ^ {+ 0.06 } _{-  0.05 }$\\
\multicolumn{1}{l|}{NR Lines}\\
$3~\mathrm{keV}$   & ${ 1.00 } ^ {+ 0.26 } _{-  0.21 }$\\
$5~\mathrm{keV}$   & ${ 1.20 } ^ {+ 0.18 } _{-  0.13 }$\\
$7~\mathrm{keV}$   & ${ 1.18 } ^ {+ 0.15 } _{-  0.12 }$\\
$10~\mathrm{keV}$   & ${ 1.10 } ^ {+ 0.12 } _{-  0.09 }$\\
$20~\mathrm{keV}$   & ${ 1.02 } ^ {+ 0.10 } _{-  0.07 }$\\
$30~\mathrm{keV}$   & ${ 1.01 } ^ {+ 0.08 } _{-  0.06 }$\\
\multicolumn{1}{l|}{SI WIMP recoils}\\
$6~\mathrm{GeV}/c^2$ & ${ 0.87 } ^ {+ 0.31 } _{-  0.25 }$\\
$10~\mathrm{GeV}/c^2$ & ${ 1.06 } ^ {+ 0.18 } _{-  0.16 }$\\
$50~\mathrm{GeV}/c^2$ & ${ 1.06 } ^ {+ 0.08 } _{-  0.07 }$\\
$100~\mathrm{GeV}/c^2$ & ${ 1.04 } ^ {+ 0.08 } _{-  0.07 }$\\
    \end{tabular}
     \caption{Table of bias and spread of the binwise upper limit with respect to the full likelihood result.}
    \label{tab:nt_validation}
\end{table}

\begin{figure}[t]
\centering
\includegraphics[width=0.98\columnwidth]{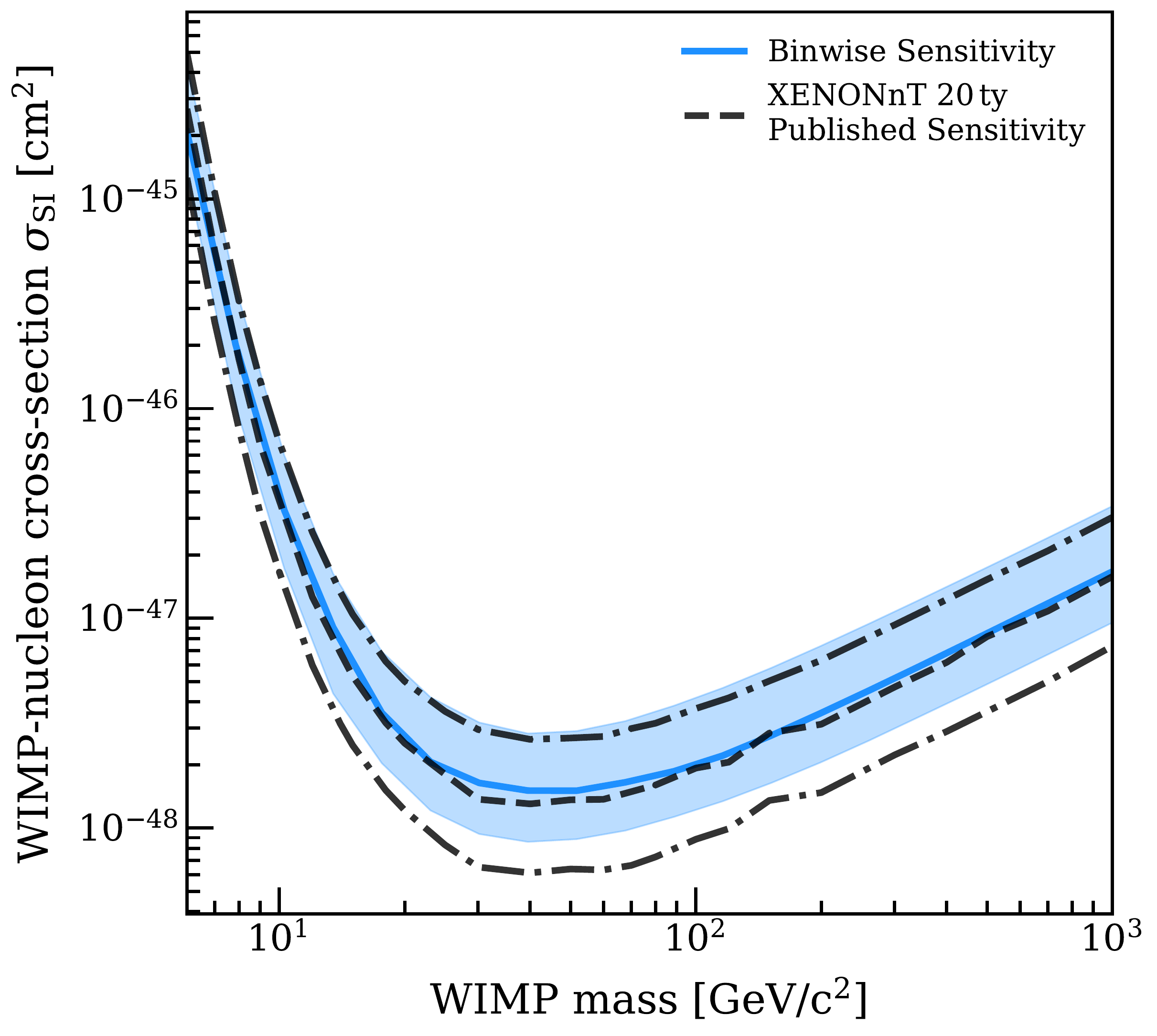}\\
\caption{\label{fig:uls_nt} Comparison between $90\%$ confidence level sensitivity bands for a projected $20$ tonne-year XENONnT search for spin-independent WIMP-nucleon interactions show good agreement between the approximate likelihood and the full likelihood. The published result using the full likelihood~\cite{xenonnt_sensitivity}  is in black. Blue line and band indicate the upper limit using the binwise approximate likelihood presented in this paper.
}
\end{figure}

We do not include a wall model in the sensitivity projections, but select a 4 tonne fiducial volume further from the detector walls than in XENON1T, to minimise this contribution. Without the addition of the wall model, we choose to model the remaining backgrounds in only the \cSone and \cStwob parameter spaces, treating their radial dependence as uniform. No background model is considered for accidental coincidence of lone S1s and S2s either. Additionally, rather than implementing a single model for CE$\nu$NS, we implement two models, one for solar neutrinos and one for the diffuse supernova neutrino background and atmospheric neutrinos.

We assume the same low-energy detection efficiency as in XENON1T, therefore the lower boundary of our analysis space in \cSone remains at 3\,PE. The higher light collection efficiency expected in XENONnT implies that more light should be detected for each scintillation photon produced, and we therefore adjust our upper boundary to 100\,PE to contain the full spin-independent WIMP recoil spectrum.

Table~\ref{tab:nt_validation} validates the recasting of the XENONnT 20 tonne-year projection, again showing good performance, and Figure~\ref{fig:uls_nt} compares the spin-independent WIMP-nucleon limits using this method and the previously published projection.

\bibliographystyle{spphys}
\bibliography{bibliography}% Produces the bibliography via BibTeX.
\end{document}